\newcommand{\kpnn}{K^0_L\rightarrow\pi^0\nu\bar{\nu}}
\newcommand{\kppp}{K^0_L\rightarrow\pi^0\pi^0\pi^0}
\newcommand{\vtx} {Z_{vtx}}
\newcommand{\pt} {P_T}

\newcommand{\red}[1]{{\color{red}\st{#1}}}

\documentclass[preprint,5p,times,twocolumn]{elsarticle} 

\usepackage{amssymb}

\usepackage{tensor}
\usepackage{xspace} 
\usepackage{graphicx}
\usepackage{dcolumn}
\usepackage{natbib}
\usepackage{color}
\usepackage{float}
\usepackage{overpic}
\usepackage{mathtools}
\usepackage{amsmath}
\usepackage{xcolor}
\usepackage{tabularx}
\usepackage{lineno}
\usepackage{soul}
\usepackage{transparent}

\journal{Nuclear Instruments and Methods in Physics Research Section A}

\begin{document}


\begin{frontmatter}



\title{ 
Suppression of Neutron Background using Deep Neural Network and Fourier Frequency Analysis at the KOTO Experiment}


\author[chicago]{Y.-C.~Tung\corref{cor1}\fnref{fn1}}
\cortext[cor1]{Corresponding author}
\fntext[fn1]{
Present address: Department of Physics, 
National Taiwan University, 
Taipei
10617,
Taiwan, Republic of China.}

\author[chicago]{J.~Li\fnref{fn2}}
\fntext[fn2]{Present address: Department of Physics and Astronomy, 
Northwestern University, Evanston, 60208, Illinois, USA.}


\author[taiwan]{Y.~B.~Hsiung}

\author[taiwan]{C.~Lin\fnref{fn5}}
\fntext[fn5]{Present address:
Enrico Fermi Institute,
University of Chicago,
Chicago, 60637,
Illinois, USA.
}

\author[osaka]{H.~Nanjo}

\author[kek]{T.~Nomura}

\author[chicago]{J.~C.~Redeker}

\author[osaka]{N.~Shimizu\fnref{fn6}}
\fntext[fn6]{Present address: Department of Physics and The
International Center for Hadron Astrophysics, Chiba
University, Chiba 263-8522, Japan.}

\author[kyoto]{S.~Shinohara\fnref{fn3}}
\fntext[fn3]{Present address: KEK, Tsukuba, Ibaraki 305-0801, Japan.}

\author[kek]{K.~Shiomi}

\author[chicago]{Y.~W.~Wah}

\author[osaka]{T.~Yamanaka}


\affiliation[chicago]{organization={Enrico Fermi Institute},
            addressline={University of Chicago}, 
            city={Chicago},
            postcode={60637}, 
            state={Illinois},
            country={USA}}
            
\affiliation[taiwan]{organization={Department of Physics},
            addressline={National Taiwan University}, 
            city={Taipei},
            postcode={10617}, 
            country={Taiwan, Republic of China}}     

\affiliation[osaka]{organization={Department of Physics},
            addressline={Osaka University, Toyonaka}, 
            city={Osaka},
            postcode={560-0043}, 
            country={Japan}}
            
\affiliation[kek]{organization={Institute of Particle and Nuclear Studies},
            addressline={High Energy Accelerator Research Organization (KEK)}, 
            city={Tsukuba},
            postcode={305-0801}, 
            state={Ibaraki},
            country={Japan}}

\affiliation[kyoto]{organization={Department of Physics},
            addressline={Kyoto University}, 
            city={Kyoto},
            postcode={606-8502}, 
            country={Japan}}

\begin{abstract}
We present two analysis techniques for distinguishing 
background events induced by neutrons
from photon signal events in the search for the rare $\kpnn$ decay at the J-PARC KOTO experiment. 
These techniques employed a
deep convolutional neural network and Fourier frequency analysis to discriminate neutrons from photons,
based on their variations in cluster shape and pulse shape,
in the electromagnetic calorimeter made of undoped CsI.
The results effectively suppressed the neutron background by 
a factor of $5.6\times10^5$,
while maintaining the efficiency of $\kpnn$ at $70\%$.

\end{abstract}

\end{frontmatter}






\section{Introduction}
\label{}

The KOTO experiment at J-PARC was designed to search for the 
rare decay of $\kpnn$, which has a theoretical branching ratio of
$\mathcal{B}_{\text{SM}}(\kpnn)=(3.00\pm0.30)\times10^{-11}$ in the standard model~\cite{buras}.
The current result on the $\kpnn$ measurement
is an experimental upper limit on the branching ratio, 
which is 
$\mathcal{B}_{\text{EXP}}(\kpnn)<3.0\times10^{-9}$ 
at the $90\%$ confidence level
set by KOTO~\cite{kpnn2015,kpnn2016}.
KOTO utilized the high-intensity $30~\mathrm{GeV}$
proton beam 
incident on a gold target to produce secondary particles, 
and the secondary neutral particles, including kaons, 
were guided to the KOTO detector 
by two sets of collimators~\cite{shimo}. 
The only visible products in the $\kpnn$ decay are
two photons from the subsequent decay of $\pi^0 \rightarrow \gamma \gamma$.
Therefore, a $\kpnn$ event is identified by two photons detected
in a Cesium Iodide crystal calorimeter (CSI)~\cite{csi}.
One of the dominant background sources in the search for $\kpnn$
was the beam-halo neutron. 
These neutrons could present a similar event signature
with two photon-like hits in the CSI. 
A halo neutron background event was caused by
a single halo neutron particle 
that interacted inside the CSI and produced two photon-like hits.
Typically, 
the first hit occurred 
near the neutron's incident point on the CSI, 
while the second hit, 
produced by the same neutron after the scattering process, 
was separated from the first hit by some distance.
These two hits in the CSI with no trace in between 
could be mistaken as 
the two isolated photon hits from $\kpnn$.
Suppressing the halo neutron background relied on distinguishing the interaction footprints of photon and neutron hits in the CSI,
which were characterized by differences in 
the incident particle's cluster shape and pulse shape in the CSI.

The CSI consisted of 2716 Cesium Iodide (CsI) crystals arranged 
in a grid format with each crystal having its own individual readout.
When a particle interacted with the CSI,
the analog pulse shape in each crystal was digitized and recorded, 
and the cluster shape was formed by grouping nearby CsI crystals 
with deposited energy by the incident particle.
Discrimination between photons and neutrons 
was based on variations in cluster shape and pulse shape.
Although the cluster shape discrimination~\cite{csi} 
and the pulse shape discrimination~\cite{yasu} 
had been previously studied at KOTO,
in this article,
we introduce two new techniques: 
using a Convolutional Neural Network (CNN)~\cite{cnn} 
to classify cluster shapes (Section~\ref{csd}) and
Fourier frequency analysis to discriminate pulse shapes (Section~\ref{psd}).
In addition, we introduce a more precise method for estimating the neutron background level in Section~\ref{performance}. 
These techniques were first introduced to the $\kpnn$ analysis of
the data obtained from the years 2016--2018~\cite{kpnn2016},
and the result further suppressed 
KOTO’s dominant background source, the halo neutron, by a factor of 26
over the previous $\kpnn$ analysis result of 2015 data~\cite{kpnn2015}.
This article provides a detailed explanation of how
neutron background events were suppressed and estimated
in the 2016--2018 data analysis,
which was not presented in the previous publications of KOTO.

\section{CsI detector}

\begin{figure*}[tb]
	\centering
	\includegraphics[width=0.9\linewidth]{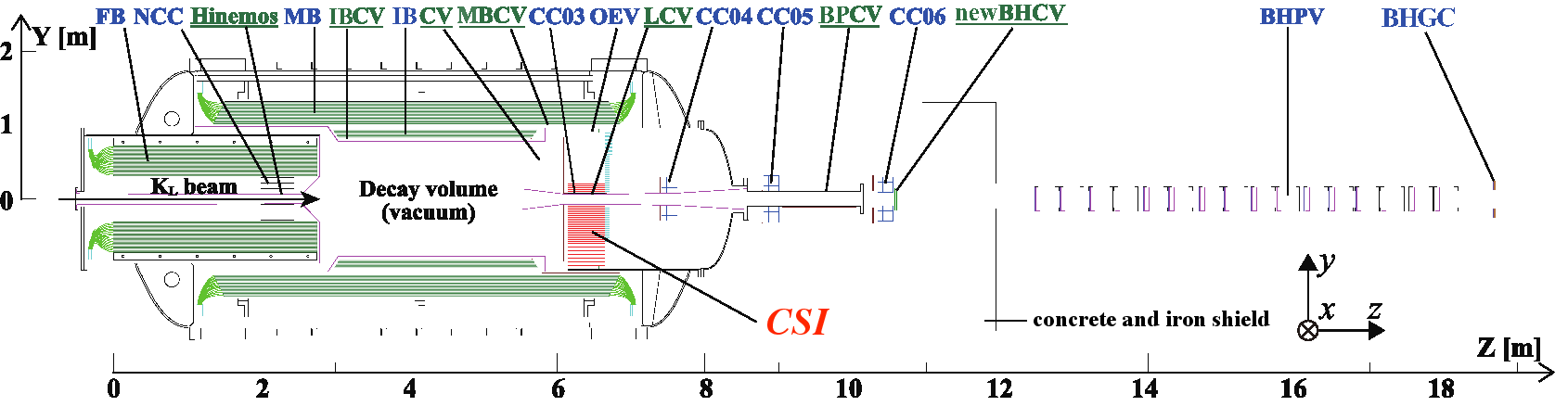}
	\caption{ 	
Cross-sectional view of the KOTO detector, with the beam entering from the left. The detector components with their names underlined represent charged-particle veto counters. The other components, except for CSI, serve as photon veto counters.
		}
	\label{fig:KOTODetector}
\end{figure*}

\begin{figure}[!h]
\begin{overpic}[width=1\linewidth]{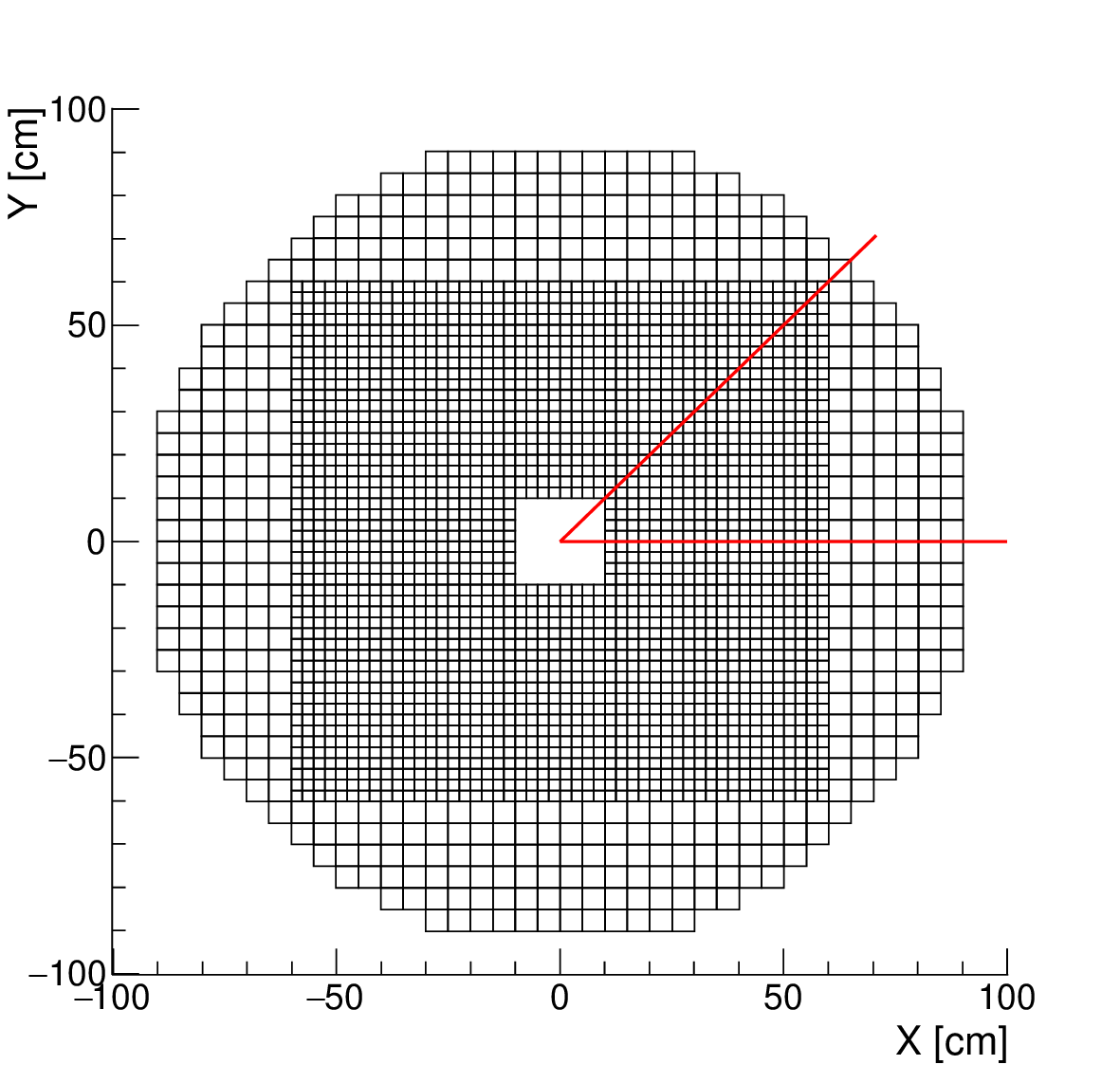}
\put(79,77){ \textcolor{red} {\large $\phi=45^{\circ}$}}
\put(91,47.5){ \textcolor{red} {\large $\phi=0^{\circ}$}}
\end{overpic}
\caption{\label{fig:csi}
Eight-fold symmetrical layout of the CSI calorimeter viewed from downstream; this allows to mirror and fold a cluster at any location on 
the CSI surface
to an equivalent location within the angular region of 
$0^{\circ} \leq \phi \leq 45^{\circ}$.
}
\end{figure}

Figure~\ref{fig:KOTODetector} 
shows a cross-sectional view of the KOTO detector,
where the coordinate origin is defined at the entrance of the detector.
The CSI is located at $z = 6.1~\mathrm{m}$, 
the downstream end of the decay volume.
It consists of 2716 undoped CsI crystals, 
covering a circular area with a radius of $90$~cm
and with a square hole for the beam to pass,
as shown in Fig.~\ref{fig:csi}.
The CSI has two different sizes of crystals. 
Crystals with a size of
$2.5\times2.5\times50~\mathrm{cm}^3$ are located in the central 
$120\times120~\mathrm{cm}^2$
square region, and others with a size of
$5.0\times5.0\times50~\mathrm{cm}^3$ 
are situated in the outer area. 
The small crystals are viewed by $3/4$~inch Hamamatsu R5364
PMTs, while the large crystals are viewed by
$1.5$~inch Hamamatsu R5330 PMTs.
The analog pulses from each PMT are digitized and recorded using 
custom-made 14-bit 125-MHz ADC modules~\cite{adc}.
For each event, 64 samples of voltages are recorded every 8~ns. 
In order to achieve a better timing resolution of the pulse, 
the analog pulse signal is reformed 
before digitization using a 10-pole Bessel filter.
This Bessel filter is used to widen and transform the PMT pulse into a Gaussian shape to increase the number of sampling points
in the pulse rising edge.
The 14-bit dynamic range of the ADC covers energy deposits from sub-MeV to 2 GeV, with an energy deposit of 1 MeV resulting in a pulse height of 8-10 ADC counts.
Further details on the CSI can be found in Ref.~\cite{csi}.

\section{Cluster shape discrimination}
\label{csd}

\subsection{Cluster shapes and data set}
\label{data_set}

\begin{figure}[h] 
	\centering
   \includegraphics[width=0.45\linewidth]{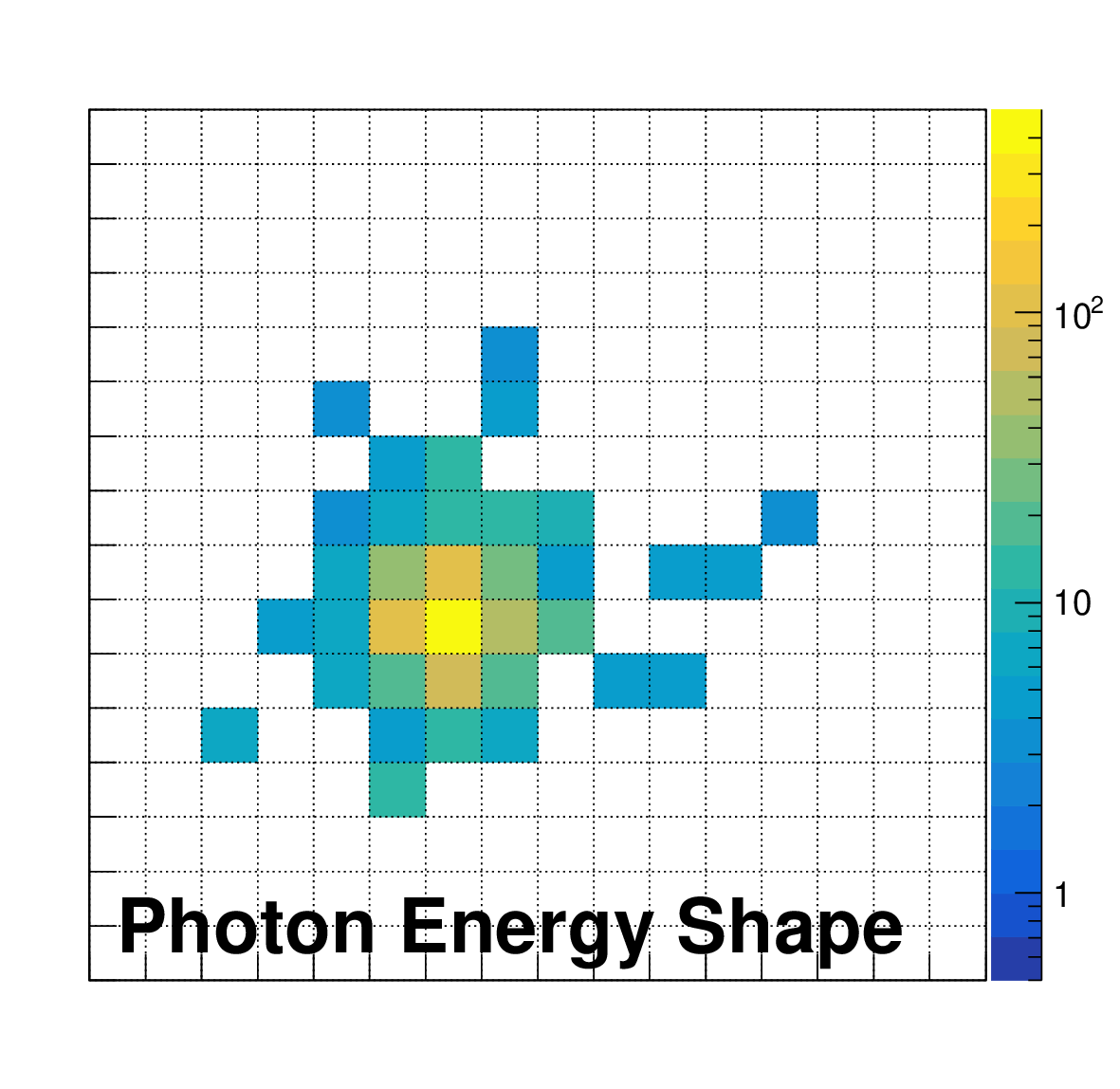}
   \includegraphics[width=0.45\linewidth]{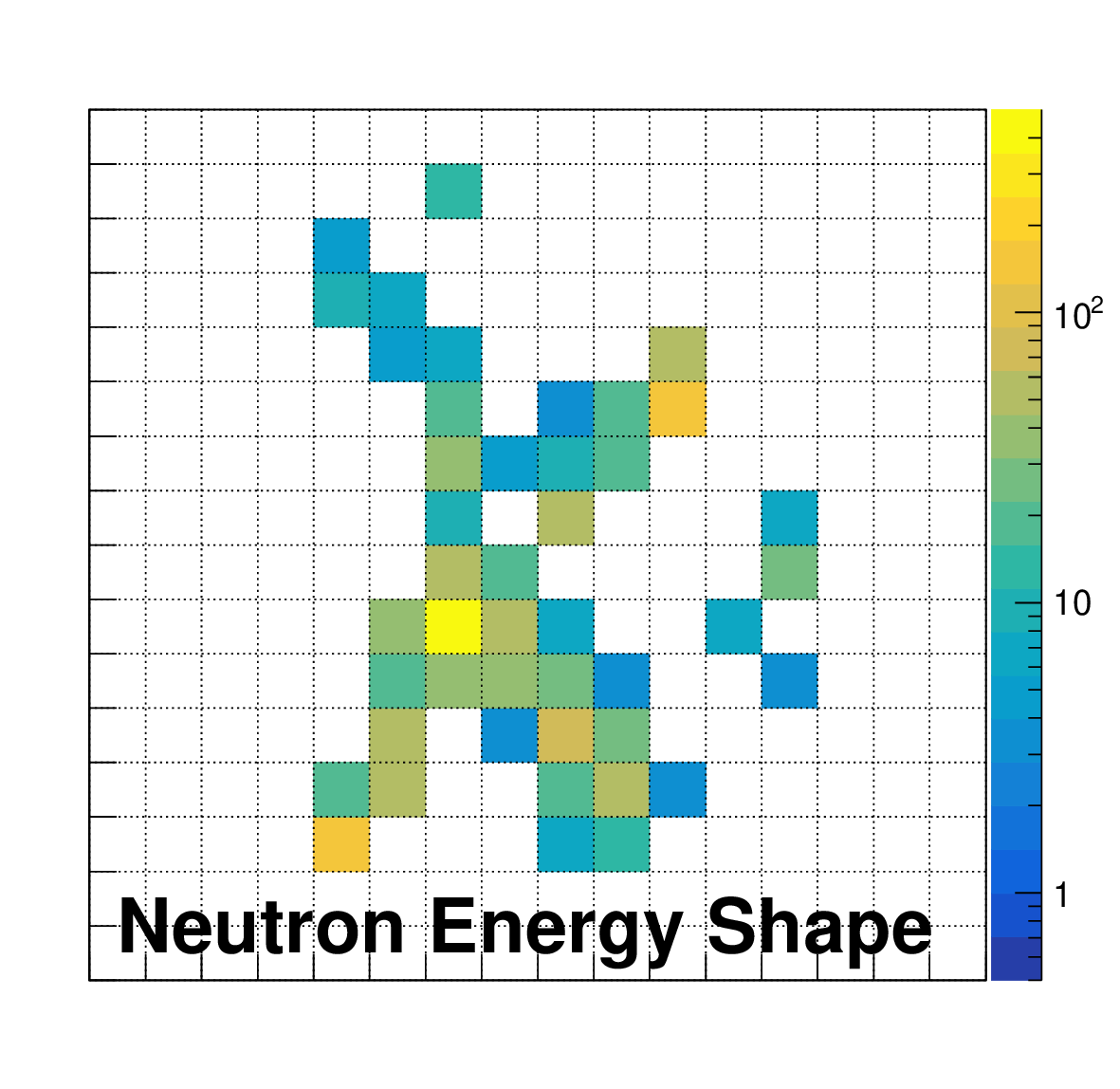}
   \includegraphics[width=0.45\linewidth]{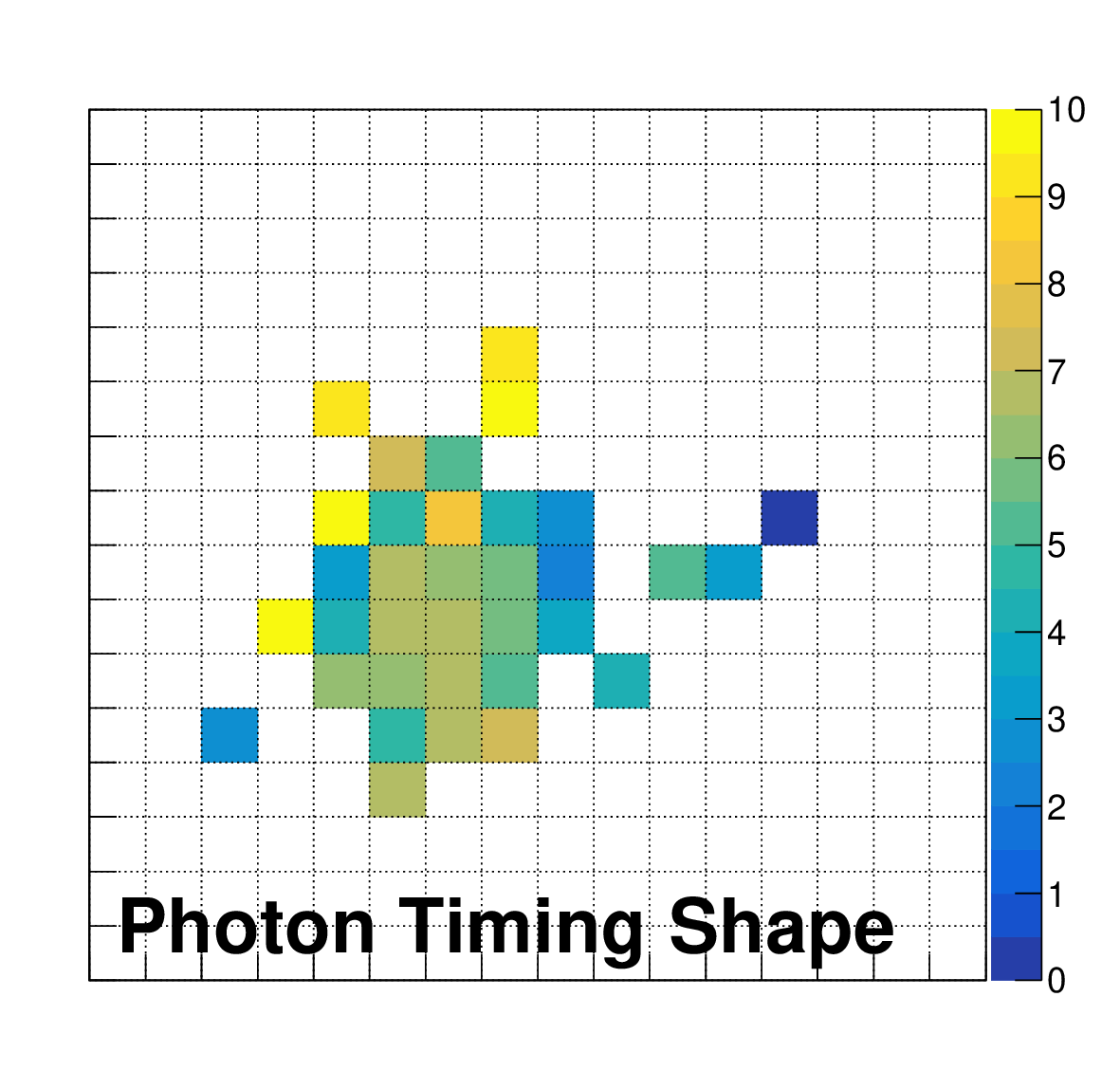}
   \includegraphics[width=0.45\linewidth]{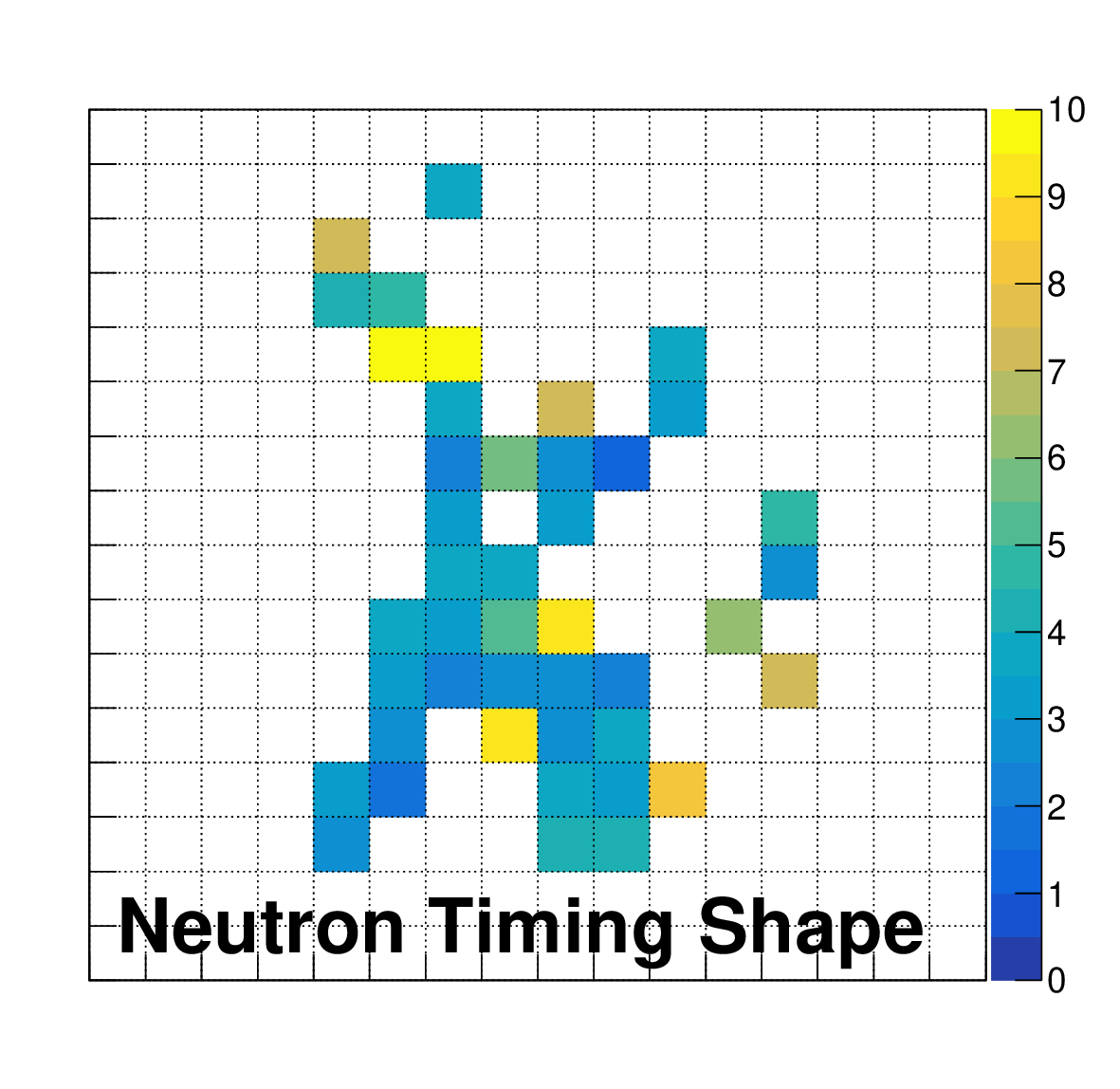}
   \caption{
   \label{fig:clusters}%
  	Example of the energy and timing shapes of 
  	photon cluster from the Monte Carlo
  	simulation (left) and neutron cluster from data (right).
	The color code represents the deposited energy in MeV and the timing in 					nanoseconds for each crystal in the cluster.
	}
\end{figure}

Cluster shape discrimination (CSD) 
is a method to distinguish
photons from neutrons, 
based on their different characteristics in the cluster shapes.
These differences arose from  
the interaction processes between
electromagnetic interactions from photons
and hadronic interactions from neutrons. 
For a pulse in a crystal,
the energy was calculated by 
integrating ADC values of the digitized waveform, and
the pulse timing was defined as 
the time of the waveform crossing half of its peak height.
These measurements of energy and timing from each crystal 
in the CSI grid formed the cluster's energy and timing 
distributions (shapes), 
which can be considered as images of particle interactions in the CSI.
Figure~\ref{fig:clusters} shows an illustration of the energy and timing shapes of photon and neutron clusters.
Typically, photon clusters produced through electromagnetic interactions
were more circular and symmetric in shapes. 
In general, the crystals closer to the incident point tend to have higher deposited energy due to the short radiation length of CSI.
If photons have finite incident angles, crystals viewing the tail of the shower tend to have earlier timing, as they have energy deposits deeper in the crystal (closer to the PMT).
On the other hand, 
neutron clusters produced through hadronic interactions 
had more asymmetrical energy and timing shapes.

In the CSD study, 
to optimize the CSD for selecting photons 
with the energy and angle spectrums from the $\kpnn$ decays,
the photon samples were obtained from $\kpnn$ Monte Carlo (MC) events, 
generated using {\scshape Geant4} simulations.
To reflect actual beam activities and electronic noise, 
the MC events were overlaid with accidental data collected 
during data-taking.
The photon samples were first selected by requiring two coincident clusters 
in the CSI and no in-time hits in other detector components. 
The decay vertex ($\vtx$) and transverse momentum ($\pt$) of $\pi^0$ were then reconstructed by assuming the $\pi^0$ decay point to lie along the beam axis and the two clusters to have the invariant mass of $\pi^0$.
The $\kpnn$ events were selected by requiring the $\vtx$ to be within the fiducial decay region of $3200$--$5000$ mm.
Additionally, 
the $\pt$ was required to be in the range of $130$--$250$ MeV/$c$,
to account for the missing momentum carried by two neutrinos.

Neutron samples were obtained from special 
neutron data-taking runs
conducted in 2016--2018. 
During these runs, an aluminum plate was placed at the detector entrance to scatter 
neutrons in the beam and enhance halo neutron events. 
The neutron data was collected 
by requiring two coincident clusters in 
the CSI and no hits in the major veto detectors. 
This sample was dominated by the events with a single neutron particle scattering within the CSI and producing two clusters. The neutron events were processed through the same reconstruction and selection procedures 
for the $\kpnn$ event candidates.


\subsection{CNN and its network architecture}
\label{cnn_arch}
\begin{figure}[h]
\begin{overpic}[width=1.\linewidth]{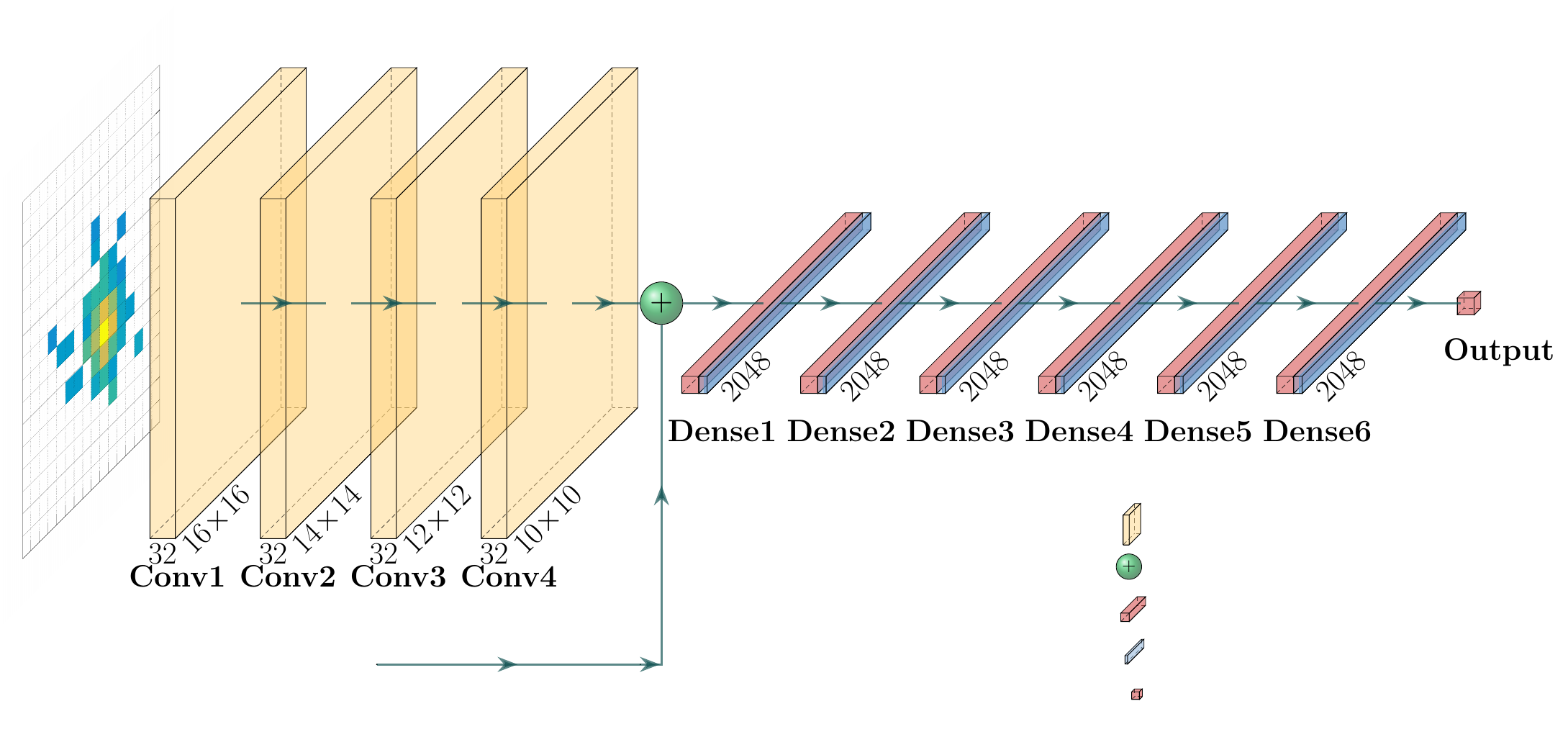}
\put(73.5,13.5){\scriptsize{:~Convolutional Layer}}
\put(73.5,10.8){\scriptsize{:~Concatenate Layer}}
\put(73.5, 8.1){\scriptsize{:~Dense Layer}}
\put(73.5, 5.4){\scriptsize{:~Dropout Layer}}
\put(73.5, 2.7){\scriptsize{:~Output Layer}}
\put(13.0, 4.1){\small{$(E,\theta,\phi)$}}
\end{overpic}
\caption{\label{fig:nnarch}
Architecture of the CSD Neural Network with 
four convolutional layers and 
six dense layers designed for 
the classification of photon and neutron cluster patterns.
}
\end{figure}
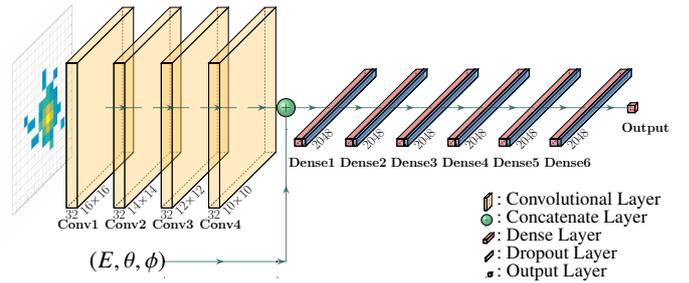

In the CSD study, a CNN was employed to differentiate between photon and neutron clusters based on their energy and timing shapes. 
The CNN is a popular deep-learning architecture 
used for image classification tasks.
The input layer of the network consisted of 
the images of photon and neutron clusters, 
which were then processed through ten hidden layers, 
including four convolutional and six dense layers,
as illustrated in Fig.~\ref{fig:nnarch}.

The convolutional layers scanned each block of $3 \times 3$ image pixels 
to identify local features of the cluster's energy and timing images 
using 32 filters.
Each filter was a tensor of $3 \times 3 \times 2$ weights and an offset bias.
The output of the fourth convolutional layer, 
along with the incident particle's energy ($E$) and direction ($\theta$, $\phi$) as additional inputs, 
were processed by six dense layers.
The dense layers were fully connected between two adjacent layers 
with 2048 neurons each to 
learn a non-linear function to produce the final output.

During network training, the neuron weights were calculated through the minimization of binary cross-entropy~\cite{dbook}, 
which served as the loss function.
This loss function provides an indication of 
the network model's performance based on the deviations 
between 
the value predicted by the model 
and the actual true value.
To prevent overtraining,
the common L2 regularization~\cite{l2}
and dropout~\cite{dropout}
techniques were applied to the network model.
The L2 regularization, with a hyperparameter of $\lambda=0.001$,
added a penalty term to the loss function at every layer of the model, 
making the neuron weights less sensitive to the training data.
A dropout layer with the dropout rate of $10\%$
was inserted between dense layers. 
During the training, 
the dropout layer randomly deactivated $10\%$ of the neurons, 
promoting the network to learn multiple representations of the data. 
These hyperparameters were fine-tuned to ensure 
the model's results were consistent between training and test samples.
Finally, the output layer produced a probability distribution 
as the CSD score, 
with values closer to 1 indicating photon-like 
and values closer to 0 indicating neutron-like.

\subsection{Training details}

The input cluster images were generated 
from the cluster shape displayed in the CSI grid.
Each pixel in the cluster image contained the energy and timing  
measured by the corresponding CsI crystal.
There were three categories of cluster images: 
the clusters with small crystals only (Type-I), 
the clusters with large crystal only (Type-II), 
and the clusters with both types of crystals (Type-III). 
In each category, 
an equal number of photon and neutron images were used 
for the network training and were divided into three separate sets: 
training, validation, and test data. 
The training set was used for the network to learn and 
adjust the weights and biases of the model. 
The validation set was used to evaluate the model's performance during the training process. 
The test set was used as a final evaluation of the performance
of trained model.
The ratio of data in the three sets was $4:1:4$.

The size of the cluster images 
in Type-I and Type-II categories were 
$16 \times 16$ and $12 \times 12$ pixels, respectively.
The Type-III cluster was treated like a
Type-I cluster by dividing each large crystal 
into four small crystals, 
each containing $1/4$ of the energy. 
However, an additional layer was added to each pixel in 
the Type-III cluster image to indicate the crystal type.
To achieve optimal performance, 
the CNN was trained separately on each of these three cluster categories.

The network was provided with additional inputs: 
the incident particle's direction $(\theta, \phi)$ 
in spherical coordinates with the origin set at the reconstructed $\vtx$. 
Here, 
$\phi$ is the azimuthal angle of the cluster 
in the CSI surface plane, 
and $\theta$ is the incident angle of the particle to the CSI surface, 
with $\theta=0^{\circ}$ pointing to the beam axis direction.
To simplify the input images and account for the eight-fold 
symmetrical layout of the CSI, 
the cluster images were mirrored and folded into the range of 
$\phi=[0^{\circ}, 45^{\circ}]$, as shown in Fig.~\ref{fig:csi}. 
To allow the network to recognize cluster patterns 
from different directions, 
the cluster images in the training and validation sets 
were duplicated by transposing the cluster image pixels 
with $(x, y) \rightarrow (y, x)$.
Moreover, 
each neutron cluster image in the training and validation samples 
was duplicated by 
randomly assigning its supplied input of incident angle $\theta$.
This data augmentation technique
enabled the network to identify neutron clusters  
from a more generalized perspective, 
regardless of the reconstructed $\vtx$.


\subsection{Training results}

\begin{figure}[h] 
   \includegraphics[width=1\linewidth]{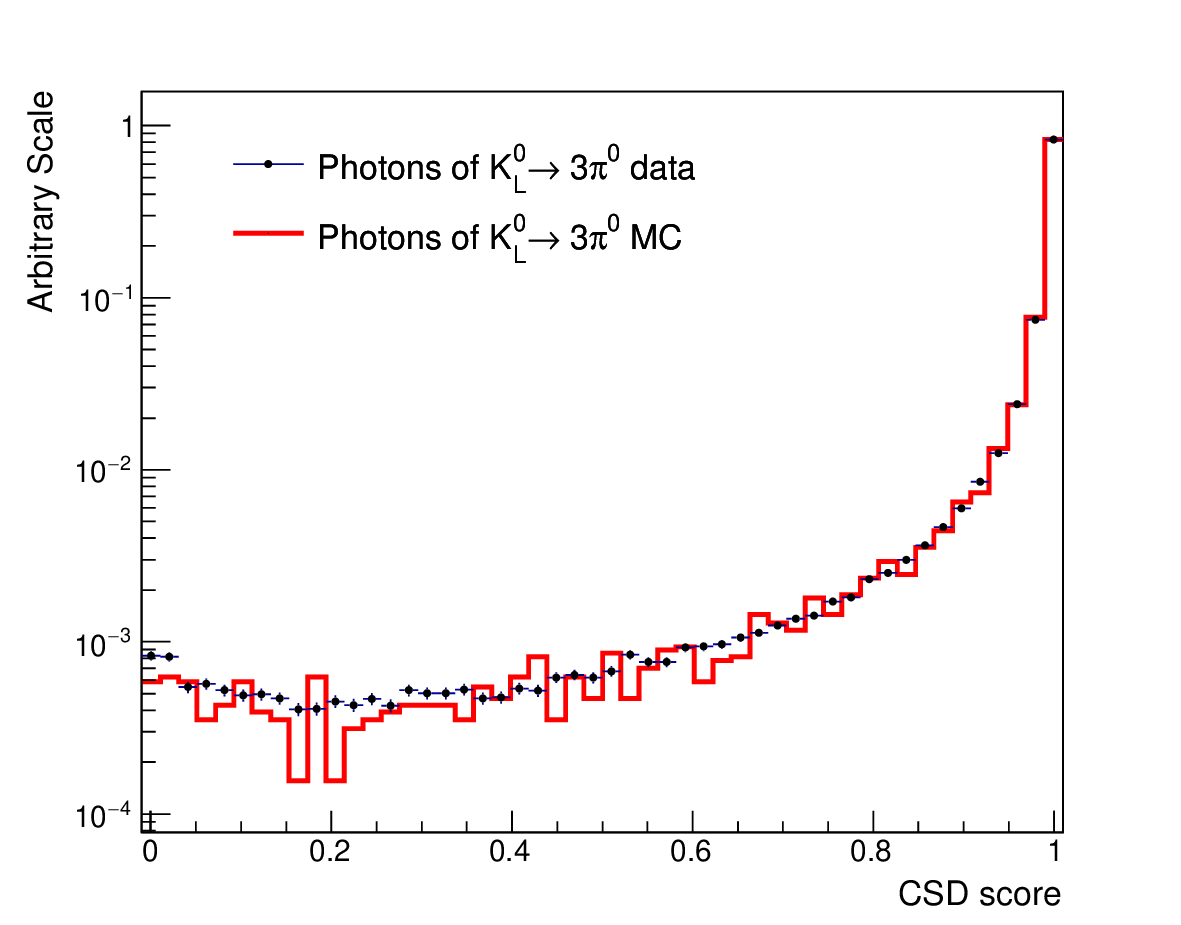}
   \caption{
   \label{fig:k3pi0}%
Distribution of the CSD score of photon clusters from $\kppp$ data (dots) and MC simulation (histogram).
	}
\end{figure}

As mentioned in Section~\ref{cnn_arch},
the training of the CSD network was optimized to prevent overfitting.
Our results indicated
that the CSD performed consistently on 
the training, validation, and test sets, 
with slightly better performance on the test set. 
This was due to the fact that 
the data augmentation was only applied to the training and validation samples.
The consistency between the data and MC simulations was verified by using the photon clusters from the $\kppp$ events. The results showed that the CSD score of the data can be accurately reproduced by the MC simulations, 
as shown in Fig.~\ref{fig:k3pi0}.
This indicates that the CSD method is reliable 
in distinguishing between photon and neutron clusters in actual data.

\begin{figure}[h] 
   \includegraphics[width=1\linewidth]{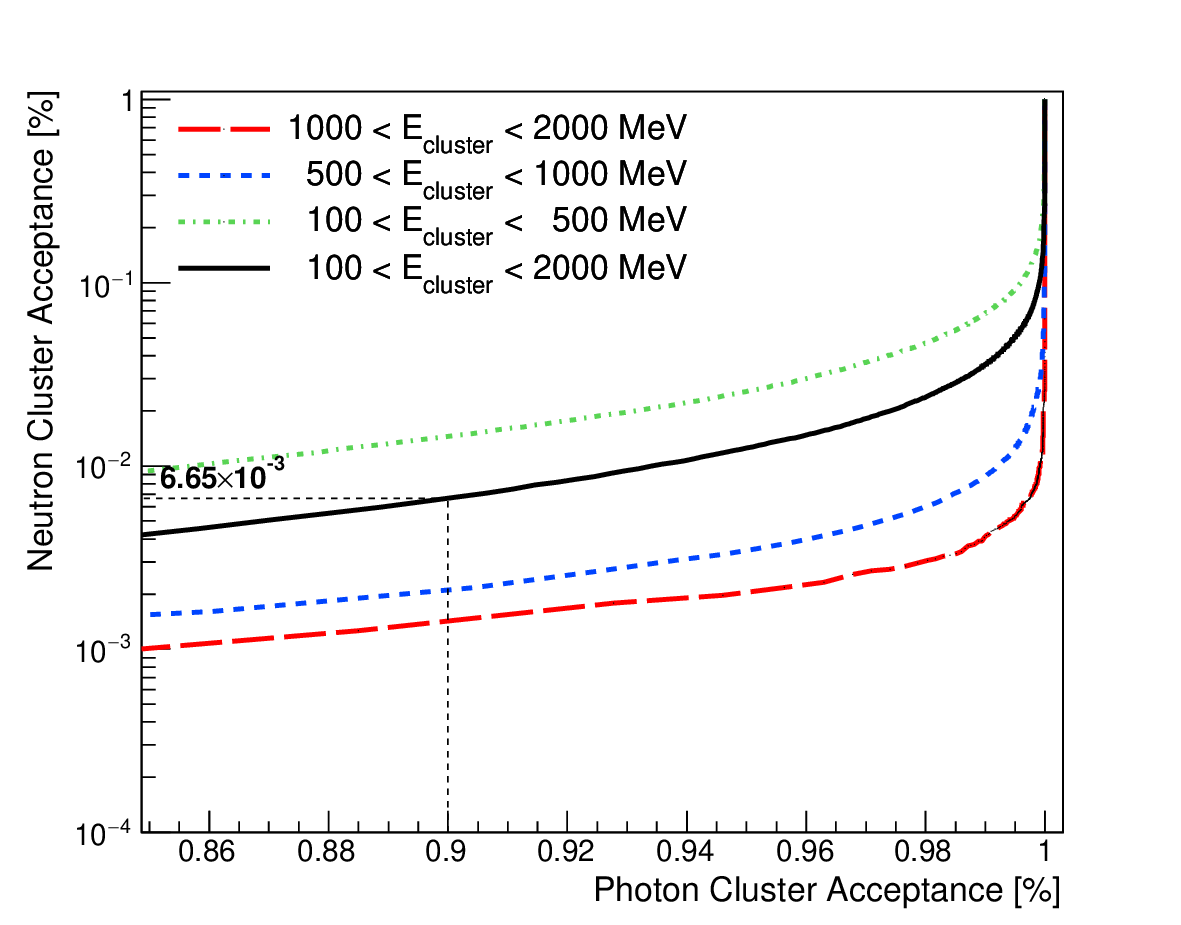}
   \includegraphics[width=1\linewidth]{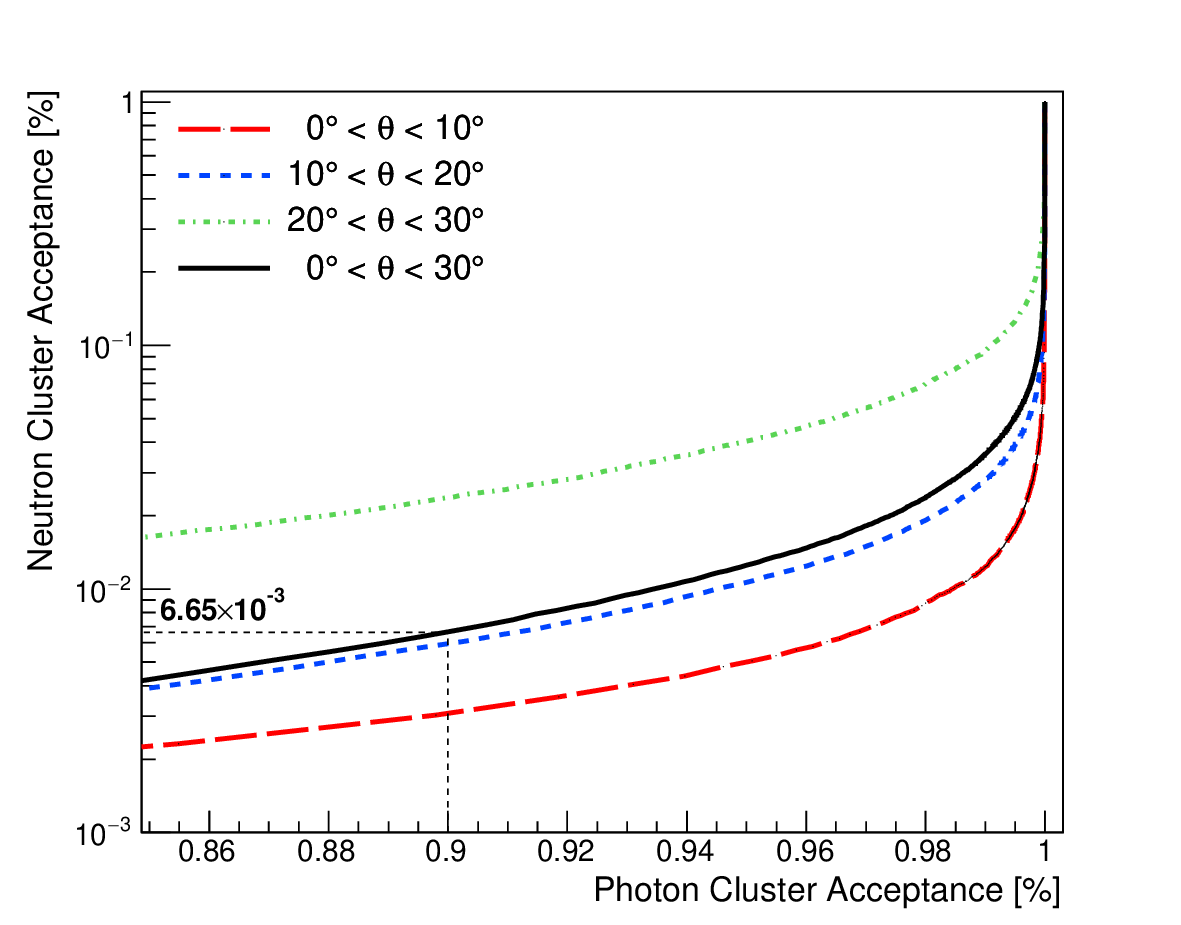}
   \caption{
   \label{fig:csd_roc}%
Performance of CSD presented as 
the acceptance of photon versus 
neutron clusters at different discrimination thresholds; 
the top figure illustrates the energy dependence, and the bottom figure displays the dependence on incident angle. 
The solid line in both figures represents the average performance 
across the energy and incident angle spectrum of 
neutron data and $\kpnn$ MC events.
	}
\end{figure}

The performance of the CSD algorithm is presented 
by the acceptance of photon clusters 
in comparison to the acceptance of neutron clusters 
at various thresholds on the CSD score, 
as shown in Fig.~\ref{fig:csd_roc}. 
In general, 
the CSD had a higher discriminating power for higher energy clusters
as larger cluster images contained more information on the cluster pattern features.
After imposing veto and kinematic selection criteria for the $\kpnn$ events,
the average energies of clusters from the neutron data samples and $\kpnn$ MC events were found to be similar, around 550~MeV.
The results based on neutron data and $\kpnn$ MC events showed that 
the CSD algorithm effectively suppressed neutron clusters by
a factor of $150$,
while maintaining a $90\%$ acceptance for photon clusters.

\section{Pulse shape discrimination}
\label{psd}

\subsection{Pulse shapes and data set}

\begin{figure}[h]
\begin{overpic}[width=1\linewidth]{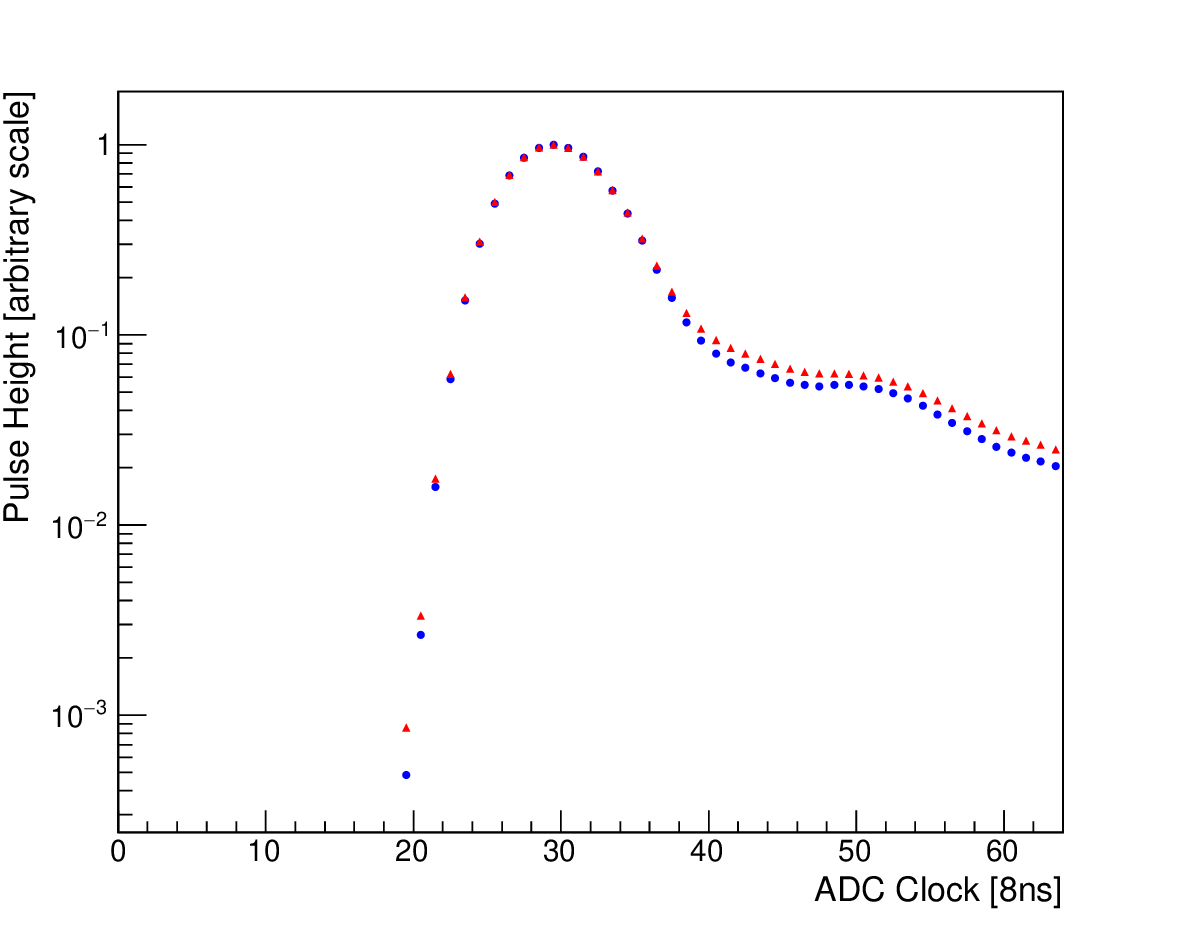}
\put(51,8.5){\includegraphics[width=0.40\linewidth]
{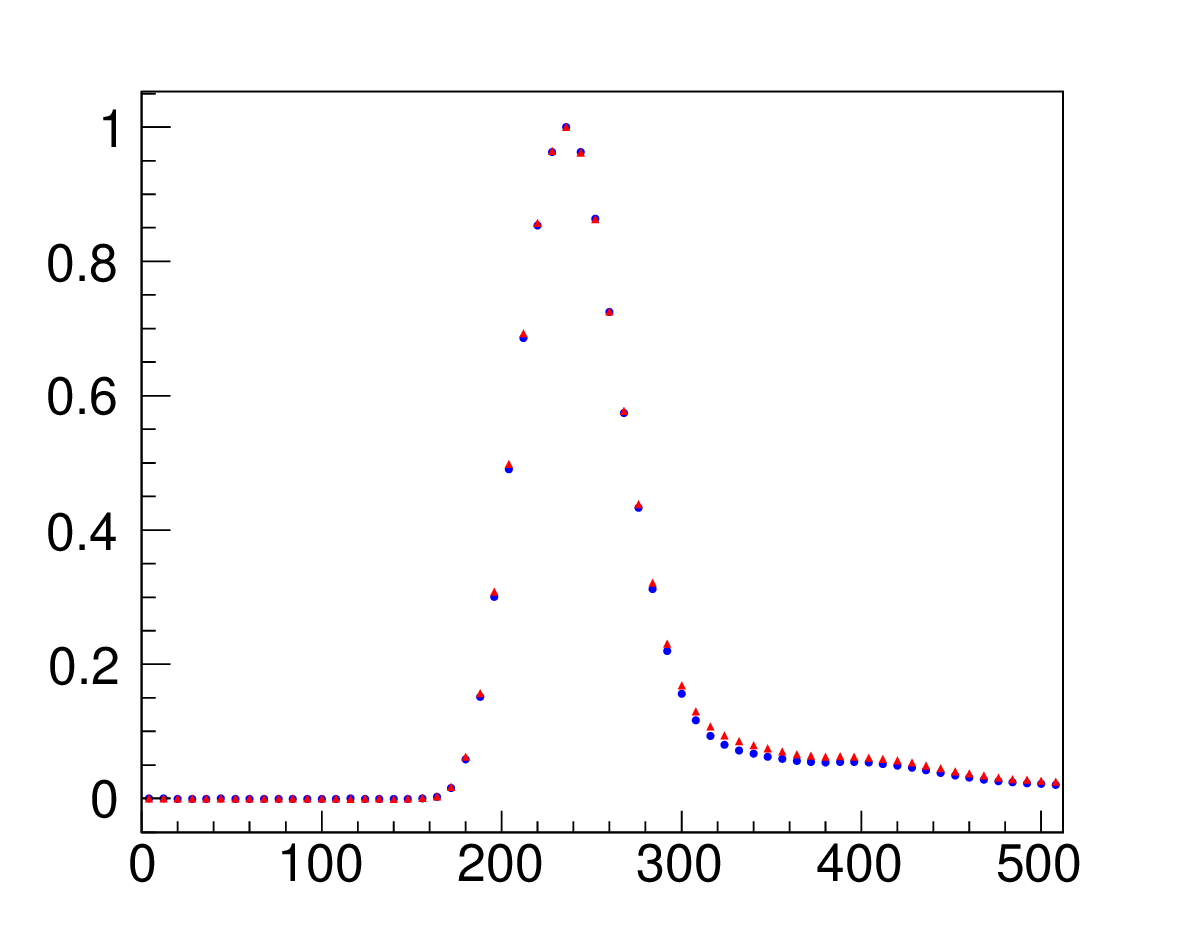}}
\put(72,33){\scriptsize Linear Scale}
\end{overpic}
\caption{\label{fig:waveform}
Average pulse shape of photon samples (blue dots) and neutron samples (red triangles) for CsI crystal ID=1013. 
}
\end{figure}

The pulse shape discrimination (PSD) is a method to 
distinguish between photon and neutron particles, 
based on their distinctive pulse shapes in the CSI.
Neutron-induced pulses through hadronic interactions
have a longer tail compared to 
those produced by photons through electromagnetic interactions, 
as shown in Fig.~\ref{fig:waveform}. 

The intrinsic differences in the detector response between photon and neutron interactions were studied using photon and neutron samples in data.
Photon samples were obtained from the data with six coincident clusters 
in the CSI and no in-time hits in major veto detectors. 
This sample was primarily dominated by $\kppp$ events. 
For the neutron samples, 
the same data set described in Section~\ref{data_set} was used.


\subsection{Discrimination method and results}

\begin{figure}[h] 
   \includegraphics[width=1\linewidth]{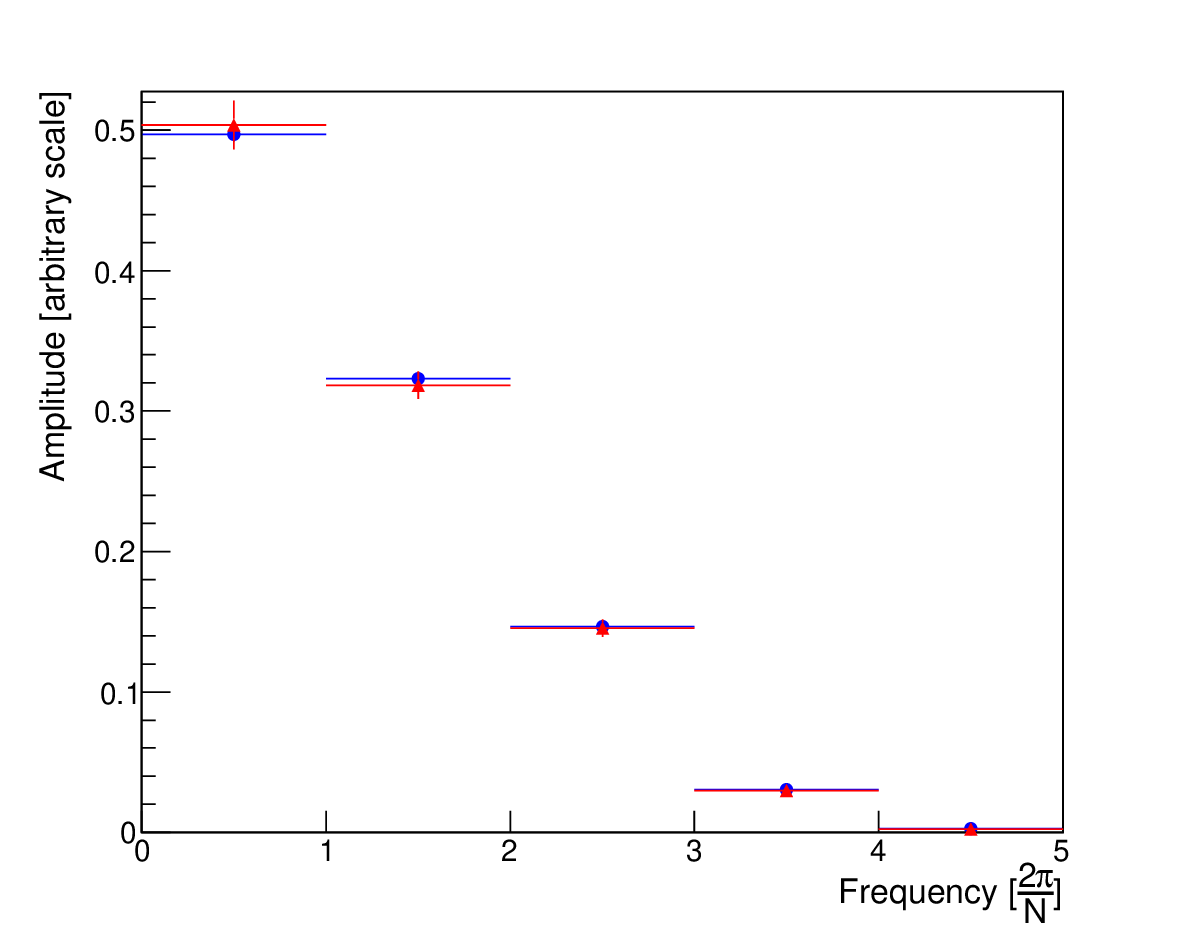}
   \caption{
   \label{fig:amp}%
Templates of photon (blue dots) and neutron (red triangles) pulses 
for CsI crystal ID=1013.
}
\end{figure}

In this study, 
the Discrete Fourier Transformation (DFT) was used to extract
the differences between the neutron and photon pulses 
in the frequency domain.
For a given CSI pulse,
the DFT was applied to the ADC values of $N_s=28$ samples: 
$
\mathcal{H}^n = 
\lbrace
\mathcal{H}^{0},\mathcal{H}^{1},...,\mathcal{H}^{N_s-1}
\rbrace
$,
where $\mathcal{H}^i$ is the ADC values of the $i^{th}$ sample.
The first sample $i=0$ was chosen
to align $\mathcal{H}^{10}$ 
with the ADC values of the pulse peak.
The DFT transformed $\mathcal{H}^n$ 
into a sequence of 28 complex numbers ($X_k$) using the equation defined as:
\begin{linenomath*}
\begin{gather}
   	X_k = \sum_{n=0}^{N^s-1} \mathcal{H}^n 
   	\text{exp}
   	\left(
   	{-\frac{i2\pi k}{N_s}n}
   	\right), 
\end{gather}
\end{linenomath*}
where 
the complex number $X_k$ encloses 
both amplitude and phase information
for the complex sinusoid at the frequency of $2\pi k/N_s$.
The tail of the CSI pulse was represented by 
the amplitude ($\mathcal{A}_k \coloneqq |X_k|$) 
of the lower frequency sinusoids.
In this analysis, 
the amplitudes of the lowest five frequency sinusoids, 
$\mathcal{A}_k = \{
\mathcal{A}_0,
\mathcal{A}_1,
\mathcal{A}_2,
\mathcal{A}_3,
\mathcal{A}_4
\}$,
were used to create templates for photons and neutrons, 
as shown in Fig.~\ref{fig:amp}.
To account for the pulse shape variations 
among crystals and energies,
templates were created for each CsI crystal and for 20 bins 
in $\mathcal{H}^{10}$ between $5.5 < \log_2(\mathcal{H}^{10}) < 14$.
Each template includes five sets of 
$\bar{\mathcal{A}_k} \pm \sigma_k$, 
where $\bar{\mathcal{A}_k}$ and $\sigma_k$ are 
the average and the standard deviation of $\mathcal{A}_k$, 
respectively.

To determine whether a given cluster is more photon-like or neutron-like, 
the likelihood of being either case 
was first calculated for each crystal contained in the cluster,
which was defined as
\begin{linenomath*}
\begin{gather}
	\mathcal{L}^{\gamma,n}_{\mathrm{crystal}} =
	\prod_{k=0}^{k<5}
	\frac{1}{\sqrt{2\pi}\sigma_k^{\gamma,n}}
	\text{exp}
	\left[
	-\frac{1}{2}
	\left(
	\frac
	{\mathcal{A}_k - \bar{\mathcal{A}}^{\gamma,n}_k}
	{\sigma^{\gamma,n}_{k}} 
	\right)^2
	\right],
\end{gather}
\end{linenomath*}
where $\mathcal{A}_k$ is the Fourier amplitudes of a crystal
in the cluster,
and $\bar{\mathcal{A}}^{\gamma,n}_k$ are the templates of  
the photon or neutron Fourier amplitudes of that crystal.
The likelihood of the cluster for being 
photon-like ($\mathcal{L}^{\gamma}_{\mathrm{cluster}}$) 
or neutron-like ($\mathcal{L}^{n}_{\mathrm{cluster}}$) 
was then calculated by multiplying the likelihood of each crystal in the cluster as
\begin{linenomath*}
\begin{gather}
		\mathcal{L}^{\gamma,n}_{\mathrm{cluster}} = 
		\prod^{N^c} 
		\mathcal{L}^{\gamma,n}_{\mathrm{crystal}},		
\end{gather}
\end{linenomath*}
where $N^c$ is the total number of crystals in the cluster.

With the photon and neutron likelihood values of a given cluster,
the final likelihood ratio $\mathcal{R}$, or PSD score, 
was calculated as
\begin{linenomath*}
\begin{gather}
   	\mathcal{R} = 
   	\frac
   	{
		\mathcal{L}^{\gamma}_{\text{cluster}}
   	}
   	{
		\mathcal{L}^{\gamma}_{\text{cluster}}
		+
		\mathcal{L}^{n}_{\text{cluster}}
   	}.
\end{gather}
\end{linenomath*}
The value of $\mathcal{R}$ is between 0 and 1, 
with $\mathcal{R}$ closer to 1 
being photon-like and $\mathcal{R}$ closer to 0 being neutron-like.

\begin{figure}[h] 
   \includegraphics[width=1\linewidth]{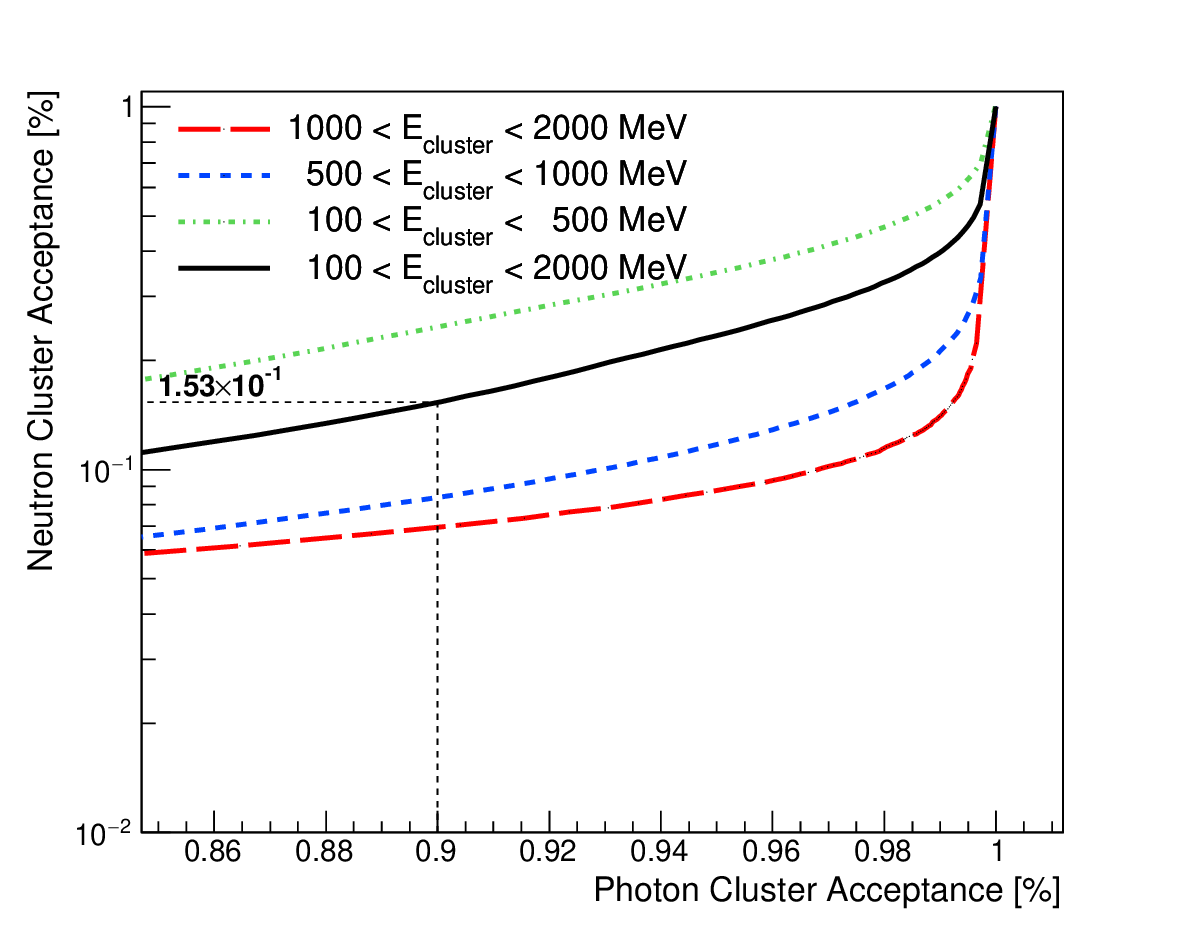}
   \caption{
   \label{fig:psd_roc}%
Performance of PSD as the acceptance of photon clusters versus 
the acceptance of neutron clusters 
at different discrimination thresholds on PSD score and
in different cluster energy regions.
The solid line represents the performance of PSD with the cluster energy spectrum of $\kpnn$.
	}
\end{figure}

The performance of PSD is
presented as the acceptance of photon clusters versus the acceptance of neutron clusters for different discrimination thresholds on the PSD score,
as shown in Fig.~\ref{fig:psd_roc}. 
The results indicate that 
the PSD is more effective in discriminating high-energy clusters. 
To evaluate the realistic performance 
in differentiating between neutron clusters and photon clusters 
in the $\kpnn$ analysis, 
the energy spectrum of photon clusters in the $\kppp$ data events was  
weighted to match that of clusters from $\kpnn$ MC events. 
It was evaluated that the PSD suppressed neutron clusters 
by a factor of $6.5$
while maintaining a $90\%$ acceptance for photon clusters from the $\kpnn$ events.
In comparison to the method described in~\cite{yasu}, 
this result with the DFT technique
performed approximately twice as effective in suppressing neutron clusters.

\section{Combined performance of CSD and PSD}
\label{performance}

The combined effectiveness of the CSD and PSD 
in suppressing neutron background events 
in the $\kpnn$ analysis was evaluated using an event-weighted method. 
This approach first calculated the survival probability ($\mathcal{W}$) 
of individual neutron clusters under the combined rejections 
of the CSD and PSD.
To take into account the energy ($E$) 
and incident angle ($\theta$) dependence 
of the CSD and PSD effectiveness, 
$\mathcal{W}$ was derived as a function of $E$ and $\theta$ 
based on a large sample of neutron clusters. 
$\mathcal{W}(E,\theta)$ 
was then used to assign event weights 
to neutron events subjected to the CSD and PSD rejections.
For a neutron event with two clusters, 
the event's survival probability was calculated 
as the product of the individual cluster's $\mathcal{W}$ values:
$\mathcal{W}_1(E_1,\theta_1) \times \mathcal{W}_2(E_2,\theta_2)$.
To estimate the remaining neutron events 
in a certain region 
after the CSD and PSD rejections,
the event-weighted method simply summed the survival probability 
of all events in that region, which gives
\begin{linenomath*}
\begin{gather}
		\mathcal{N}^{\text{Est}} 
		= \sum_i 
		\mathcal{W}^i_1(E_1,\theta_1) \times \mathcal{W}^i_2(E_2,\theta_2),
\end{gather} 
\end{linenomath*}
where $i$ counts the number of events in that region 
before the CSD and PSD rejections,
and the $\mathcal{N}^{\text{Est}}$ was the estimated number of neutron events 
that may remain after the CSD and PSD rejections.

\begin{figure}[h] 
	\centering
   \includegraphics[width=1.\linewidth]{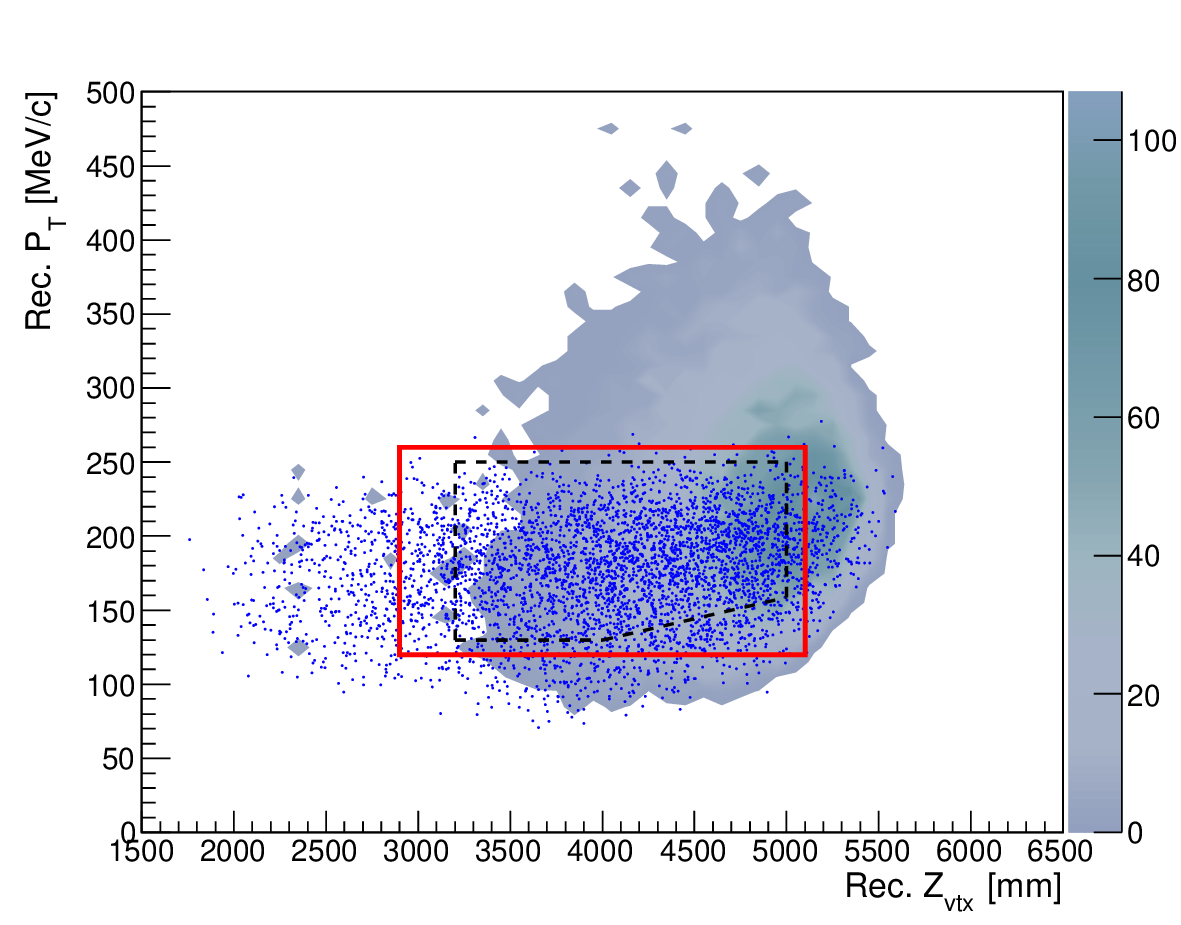}

   \caption{
   \label{fig:box}%
$\vtx~vs.~\pt$ distributions of the $\kpnn$ MC events (dots) and
neutron data events (contour). 
The dash-lined region indicates the signal region of $\kpnn$ used in
2016--2018 data analysis, 
and the solid-lined rectangular box indicates 
the Wider Signal Box (WSB) region used 
for estimating 
the performance of the combined CSD and PSD discriminations. 
}
\end{figure}

In this study, 
the neutron events in the Wide Signal Box (WSB) region
were used for evaluating the effectiveness 
of the CSD and PSD in suppressing neutron background events.
The WSB region was defined as 
$2900<\vtx<5100$ mm and $120 < \pt < 260$ MeV/$c$, 
as indicated in Fig.~\ref{fig:box}.

After imposing the $\kpnn$ event selection criteria
without CSD and PSD
to the neutron data,
there were 5973 neutron events ($\mathcal{N}^{\mathrm{Total}}$)
in the WSB region.
Based on these events, 
the estimated ($\mathcal{N}^{\mathrm{Est}}$)
and observed ($\mathcal{N}^{\mathrm{Obs}}$)
number of neutron events, 
and the $\kpnn$ efficiency ($\mathcal{E}^{\mathrm{Sig}}$)
after further imposing the CSD and PSD rejections with
different thresholds 
are summarized in Table~\ref{table:results}.
With loose thresholds on the CSD and PSD scores, 
the results indicate that the 
number of observed neutron events ($\mathcal{N}^{\mathrm{Obs}}$) 
could be accurately predicted by $\mathcal{N}^{\mathrm{Est}}$, 
demonstrating the reliability of the event-weighted method.
In the $\kpnn$ analysis of 2016--2018 data, 
the thresholds on the CSD and PSD scores were set at 
0.985 and 0.5, respectively. 
Under these thresholds, 
the event-weighted method predicted $0.0106\pm0.0002$ 
remaining neutron events in the WSB region, 
while no events were actually observed.
The acceptance of neutron events against the CSD and PSD rejections
($\mathcal{R}^{\mathrm{CSD+PSD}}$) 
was then calculated to be
\begin{linenomath*}
\begin{gather*}
		\nonumber
		\mathcal{R}^{\mathrm{CSD+PSD}}
		=
		\frac
		{\mathcal{N}^{\mathrm{Est}}}
		{\mathcal{N}^{\mathrm{Total}}}
		= (1.77\pm0.03) \times 10^{-6},	
\end{gather*} 
\end{linenomath*}
which corresponds to a suppression factor of
$5.6\times10^5$ on the neutron background events.
The efficiency of detecting $\kpnn$ under the same thresholds was determined
to be $69.9\%$.

{
\centering	
\begin{table}[H]
\caption{ 
Summary of the $\kpnn$ efficiency 
($\mathcal{E}^{\mathrm{Sig}}$)
and 
the number of observed ($\mathcal{N}^{\mathrm{Obs}}$) 
and predicted ($\mathcal{N}^{\mathrm{Est}}$) neutron background events
with different thresholds on the CSD and PSD scores. 
The asterisk marks the thresholds used in the $\kpnn$ analysis 
of 2016--2018 data.
}
\begin{tabularx}{\linewidth}{XXccc}
\hline\hline
CSD$^{Thres}$ & PSD$^{Thres}$ & $\mathcal{E}^{\mathrm{Sig}}$ ($\%$) 
& $\mathcal{N}^{\mathrm{Obs}}$
& $\mathcal{N}^{\mathrm{Est}}$  \\
\hline
\hline
- 		& - 		& $100.0$ 		& 5973 & -\\
0.0001 	& 0.0001 	& $99.8$ 		& 528 	& $528.2 \pm 10.8$\\
0.0005 	& 0.0005 	& $99.8$ 		& 161 	& $165.1 \pm 3.4$\\
0.001 	& 0.001 	& $99.8$ 		& 105 	& $102.6 \pm 2.2$\\
0.005 	& 0.005 	& $99.6$ 		& 34 	& $35.8 \pm 0.8$\\
0.01 	& 0.01 		& $99.5$ 		& 22 	& $21.8 \pm 0.5$\\
0.05 	& 0.05 		& $98.8$ 		& 5 	& $5.9 \pm 0.1$\\
0.1 	& 0.1 		& $98.0$		& 4 	& $3.1 \pm 0.1$\\
0.5 	& $0.5^{*}$ & $87.7$		& 0 	& $0.29 \pm 0.01$\\
$0.985^{*}$ & $0.5^{*}$ & $69.9$ & 0 	& $0.0106 \pm 0.0002$\\
\hline\hline
\end{tabularx}
\label{table:results}
\end{table}
}

\section{Conclusion}

We developed two analysis techniques to distinguish 
between photon and neutron events in the undoped CsI calorimeter of KOTO.
These methods were based on the distinct characteristics 
in their cluster shapes displayed on the CSI grid
and pulse shapes in each CsI crystal.
We employed a convolutional neural network to classify 
the cluster shapes and 
Fourier frequency analysis to differentiate between the waveform shapes.
The performance of discrimination was estimated through
an event-weight method. 
As a result, 
we suppressed the neutron background events by 
a factor of $5.6\times10^{5}$, 
while maintaining the acceptance of $\kpnn$ at $69.9\%$.

\section*{Acknowledgement}
We would like to express our gratitude to all members of the J-PARC Accelerator and Hadron Experimental Facility groups for their support. We also thank the KEK Computing Research Center for KEKCC, the National Institute of Information for SINET4, and the University of Chicago Computational Institute for the GPU farm. This material is based upon work supported by the Ministry of Education, Culture, Sports, Science, and Technology (MEXT) of Japan and the Japan Society for the Promotion of Science (JSPS) under KAKENHI Grant Number JP16H06343 and through the Japan-U.S. Cooperative Research Program in High Energy Physics; the U.S. Department of Energy, Office of Science, Office of High Energy Physics, under Awards No. DE-SC0009798; 
the National Science and Technology Council (NSTC) 
and Ministry of Education (MOE) in Taiwan, 
under Grants No. NSTC-111-2112-M-002-036, 
NSTC-110-2112-M-002-020, MOE-111L894805, and MOE-111L104079
through National Taiwan University.
We would also like to thank K. Kotera
for the critical reading and suggestions for this article.





\end{document}